\documentclass[twoside]{articlek}

\renewcommand{\refname}{REFERENCES}

\def\be{\begin{equation}}
\def\ee{\end{equation}}
\def\BibTeX{{\rm B\kern-.05em{\sc i\kern-.025em b}\kern-.08em
            T\kern-.1667em\lower.7ex\hbox{E}\kern-.125emX}}

\usepackage{graphicx}
\begin{document}
\sloppy
\twocolumn[{
%\vspace*{1.7cm}   % for the title  page only
%\begin{center}
{\large\bf LHC AND THE STRONGLY-INTERACTING EXTENSIONS OF THE STANDARD MODEL}\\

{\small M. Gintner, gintner@fyzika.uniza.sk, Physics Department, 
University of \v{Z}ilina, \v{Z}ilina, Slovakia, and 
the Institute of Experimental and Applied Physics, Czech Technical University,
Prague, Czech Republic}\\

%\end{center}
%\vspace*{1ex}

%{\bf ABSTRACT.} These are the guidelines for preparation of
%manuscripts of the contributions to the proceedings of the 17th
%Conference of Slovak Physicists, held on September 16 -
%19, 2009.\\
}]
%---------------------------------------------------------------------------
\section{A NEW 125-GEV BOSON}
\label{sec:New125GeVBoson}

On July 4, 2012, the representatives of two major LHC experiments,
Fabiola Gianotti of the ATLAS collaboration and Joseph Incandela of
the CMS, presented at the CERN's Main Auditorium discovery of 
a new particle~\cite{July4seminar}. The discovery was an outcome of the long-lasting
effort to find or exclude the Standard model (SM) Higgs boson, 
a possible participant of physics responsible for electroweak
symmetry breaking (ESB). The ATLAS collaboration
announced the $5.9\sigma$ signal of a boson of the mass of
$126.0\pm 0.4\mathrm{(stat.)}\pm 0.4\mathrm{(sys.)}$~GeV.
The CMS collaboration observed similar signal of a boson
with the mass of $125.3\pm 0.4\mathrm{(stat.)}\pm 0.5\mathrm{(sys.)}$.

The crossing of the magical ``five sigma'' deviation which
in the high energy physics is traditionally considered as 
a ``must'' for claiming a discovery has been a result of combining
events from more than two decay channels. Nevertheless, two of the channels
have contributed far the most to the achievements of both experiments.
The crucial channels have been
$h\rightarrow\gamma\gamma$ and $h\rightarrow ZZ^\ast\rightarrow\ell\ell\ell\ell$,
where $h$ stands for the newly discovered 125-GeV boson;
do not read $h$ as ``Higgs boson'', though. The question if the discoveree
is a/the Higgs boson will be discussed below. The discoveries
of both collaborations have been published in~\cite{125GeVbosonATLAS,125GeVbosonCMS}.

Shortly after the announcement, the CDF and D0 collaborations of the Tevatron
$p\bar{p}$ collider in Fermilab, USA, published their findings~\cite{125GeVbosonTevatron},
potentially related to the new discovery. In the Tevatron Run II, 
the CDF and D0 observed together the $3.1\sigma$ excess of events in
the would-be $h\rightarrow b\bar{b}$ channel over the invariant mass
interval of $(120;135)$~GeV.

While the discovery and all accompanying observations resulted from
the dedicated effort to find or exclude the SM Higgs boson the current
data does not prove that the new 125-GeV particle really is the SM Higgs.
Let us try to summarize what can and cannot be said about the new particle at the very 
moment. We can claim the discovery of a new particle of the mass of about
125~GeV. It is not clear whether it is elementary or composite.
Based on its decay products and the conservation laws
the particle is electrically neutral and cannot be a fermion.
Thus, it is a boson. Since it decays to two photons the Landau-Yang 
theorem~\cite{Landau,Yang}
excludes also the spin 1. The new particle is color-neutral, i.e.\ it does
not ``feel'' the strong nuclear force. Finally, based on the number of
the observed events one can deduce that the coupling of the new particle
to $Z$ boson is two orders of magnitude stronger than its coupling to
photon.

%%%%%%%%%%%%%%%%%%%%%%%%%%%%%%%%%%%%%%%%%%%%%%%%%%%%%%%%%%%%%%%%%%%%%%%%%
\section{IS IT A HIGGS?}
\label{sec:IsItHiggs}

The Higgs particle is a possible byproduct of the mechanism
responsible for ESB. 

The obvious fact that the masses of many of the known elementary particles, the $W$ and
$Z$ bosons at the first place, have non-zero values seems to break the gauge
symmetry of the electroweak Lagrangian. The gauge symmetry
represents a very useful and well verified guiding principle in building interaction 
structure of the SM Lagrangian. However, the straightforward introduction of the non-zero mass 
terms into the Lagrangian breaks the symmetry. Fortunately, Peter Higgs and
others~\cite{Higgs,EnglertBrout,GuralnikHagenKibble} 
found the solution to the problem. Particle fields can obtain
non-zero masses without sacrificing the Lagrangian's gauge symmetry if the symmetry 
of the theory's vacuum is properly lower than the gauge symmetry. This
concept is known as \textit{spontaneous symmetry breaking} (SSB). 
Thus, through SSB we can reconcile the apparent fact of ESB with the observed symmetry
of the electroweak interactions.
It is certainly encouraging that we know physical
systems in nature where SSB is actually at work. The classical examples are
represented by the phenomenon of superconductivity and by physics of hadrons.

The SM contains a particular formulation of the ESB mechanism. It is based
on introducing a new $SU(2)_L$ complex scalar doublet field with non-zero
vacuum expectation value, $v\approx 250$~GeV. Out of the four real scalar fields 
of the doublet three fields are unphysical and the fourth one is a real
particle called the SM Higgs boson.
However, this is not the only possibility how to realize ESB spontaneously.
It is the simplest possibility, in a sense, and thus, naturally, the first
candidate for investigation. Many alternative mechanisms have been formulated,
though. The numbers of Higgs-like particles
predicted by them range from zero to several.
We can say that the question of the mechanism responsible for ESB has become a centerpiece of
all speculations about physics beyond the SM.

When we turn to the recent 125-GeV boson discovery we should ask if the new particle 
has any relation
to the ESB mechanism. 
The particle should be called a Higgs boson of some sort only if the answer is yes.

Since the $W$ and $Z$ boson obtain their masses through the ESB mechanism
we expect that the coupling of a Higgs boson to $W$ and $Z$ will be 
significantly stronger than the coupling to the massless photon. If, on the other hand,
the new particle $h$ had no connection to ESB we would expect 
$g(h\gamma\gamma)\approx g(hZZ)\approx g(hWW)$. 
The existing data supports the claim that the new particle \textit{is} related
to the mechanism of ESB.

The spontaneous ESB can provide masses for fermions as well. As in the case
of the electroweak bosons $W$ and $Z$, the couplings of a Higgs particle 
to fermions should be proportional to fermion masses. Is it the case?
Unfortunately, there is not enough data at this moment to make clear conclusion
about this question and we have to wait till the end of the 2012 LHC run,
at least.

We can proceed a one step further in questioning the nature of the newly
discovered boson: what is the data support for the boson being 
the SM Higgs boson? Should it be answered in a single sentence, 
then the data roughly resembles 
the SM Higgs boson and cannot exclude it. 

To distinguish the SM Higgs boson nature of the observed particle
we have to compare theory with experiment. Since we know the mass
of the SM Higgs boson candidate, 125~GeV, we can calculate its total decay
width, $\Gamma_{\mathrm{tot}}^{\mathrm{SM Higgs}} = 4.2$~MeV, and 
the branching ratios of individual decay channels. They can be read
off of Fig.~1. There we can see that the dominant decay channel of
the 125-GeV SM Higgs boson is $h\rightarrow b\bar{b}$.
%%%%%%%%%%%%%%%%%%%%%%%%%%%%%%%%%%%%%%%%%%%%%%%%%%%%%%%%%%%%%%%%%%%%%%%%%
\begin{figure} [h,t]                     %instead of \begin{figure}[t]
\begin{center}                        %instead of \begin{center}
\includegraphics[width=80mm]{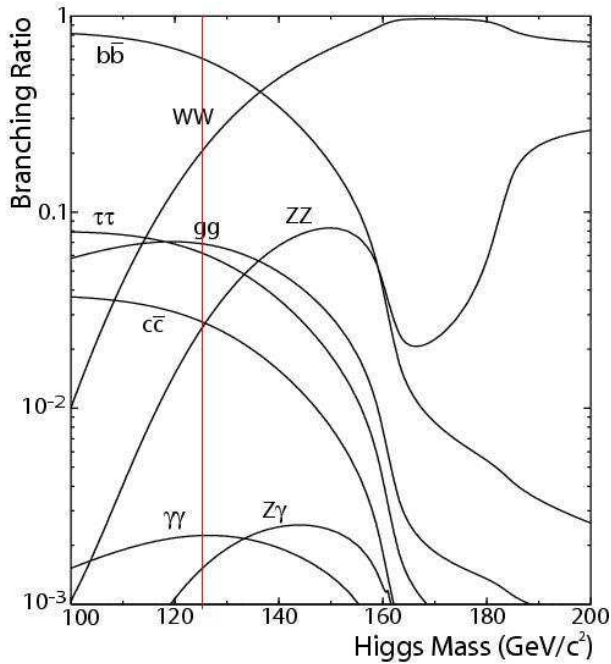}   % for LaTeX 2e
\end{center}                         %instead of \end{center}
\vspace{-2mm} \caption{The SM Higgs boson branching ratios as
                       functions of the Higgs boson mass. The mass
                       of 125~GeV is indicated by the vertical line.}
\end{figure}                        %instead of \end{figure}
%%%%%%%%%%%%%%%%%%%%%%%%%%%%%%%%%%%%%%%%%%%%%%%%%%%%%%%%%%%%%%%%%%%%%%%%%
On the other hand, the branching ratios of $h\rightarrow ZZ^\ast$ and 
$h\rightarrow\gamma\gamma$, are the one and two orders of
magnitude, respectively, smaller. Despite that, these were
the major discovery channels of the 125-GeV boson. This is because
the more dominant channels are plagued by the huge backgrounds of
the LHC $pp$ collisions. Thus, to discover the Higgs particle 
of this particular mass was difficult. On the other hand, this particular
mass provides us with a large number of decay channels to study once
the difficulties are overcome. 

Probing the SM Higgs boson nature of the 125-GeV boson can be split into
two steps. First, the new 125-GeV particle should be discovered in all
SM Higgs boson decay channels. That would confirm the existence of all
the decay channels. In Table~1 we show the decay channels of 
the greatest statistical significance currently observed.
%%%%%%%%%%%%%%%%%%%%%%%%%%%%%%%%%%%%%%%%%%%%%%%%%%%%%%%%%%%%%%%%%%%%%%%%%
\begin{table}[ht]
\begin{center}
 \caption{The most pronounced statistical significances of the 125-GeV 
 boson signals observed in particular decay channels at the ATLAS,
 CMS, and CDF+D0 detectors~\cite{125GeVbosonATLAS,125GeVbosonCMS,125GeVbosonTevatron}.}
\begin{tabular}{|c|c|c|c|}
\hline
channel & ATLAS & CMS & Tevatron \\
\hline\hline
$\gamma\gamma$ & $4.5\sigma$ & $4.1\sigma$ & - \\
\hline
$ZZ^\ast$ & $3.6\sigma$ & $3.2\sigma$ & - \\
\hline
$WW^\ast$ & $2.8\sigma$ & $1.6\sigma$ & - \\
\hline
$b\bar{b}$ & - & - & $3.1\sigma$ \\
\hline
\end{tabular}
\end{center}
\end{table}
%%%%%%%%%%%%%%%%%%%%%%%%%%%%%%%%%%%%%%%%%%%%%%%%%%%%%%%%%%%%%%%%%%%%%%%%%
There is a good chance that the question of the existence of the channels listed 
in Table~1 and $h\rightarrow\tau\tau$ will be settled by 
the full 2012 LHC data. Unfortunately, the LHC is not capable to detect
the $h\rightarrow c\bar{c}$ channel.

The second step is intimately related to the first one. It includes 
comparing the relative signal strengths of the individual decay channels
with the SM predictions. The relative signal strength can be defined as
\begin{equation}
 \mu \equiv \frac{\left(\sigma_\mathrm{prod}\times\mathrm{BR}\right)_{\mathrm{observed}}}
                 {\left(\sigma_\mathrm{prod}\times\mathrm{BR}\right)_{\mathrm{SM}}},
\end{equation}
where $\sigma_\mathrm{prod}$ is the Higgs boson production cross section
and $\mathrm{BR}$ is the branching ratio of the Higgs boson to a particular
channel. The ``no decay'' in a particular channel would result in $\mu=0$,
while the signal strength equal to that of the SM would 
result in $\mu=1$.

In Fig.~2, there are signal strengths of individual
decay channels observed at the ATLAS detector depicted.
%%%%%%%%%%%%%%%%%%%%%%%%%%%%%%%%%%%%%%%%%%%%%%%%%%%%%%%%%%%%%%%%%%%%%%%%%
\begin{figure} [h,t]                     %instead of \begin{figure}[t]
\begin{center}                        %instead of \begin{center}
\includegraphics[width=80mm]{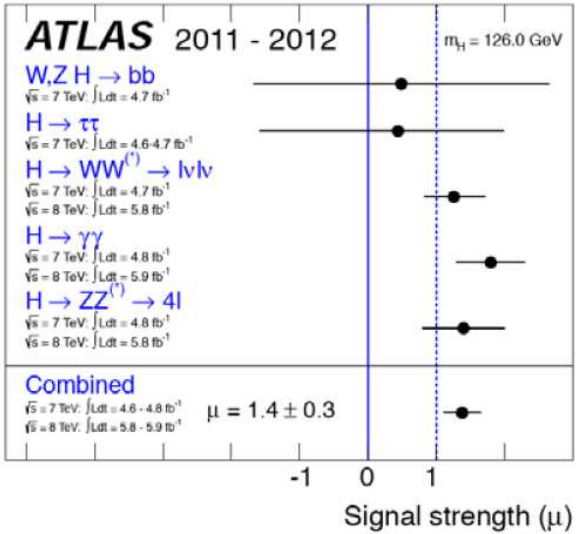}   % for LaTeX 2e
\end{center}                         %instead of \end{center}
\vspace{-2mm} \caption{The signal strengths $\mu$ of the 125-GeV boson decay channels
observed at the ATLAS detector~\cite{125GeVbosonATLAS}.}
\end{figure}                        %instead of \end{figure}
%%%%%%%%%%%%%%%%%%%%%%%%%%%%%%%%%%%%%%%%%%%%%%%%%%%%%%%%%%%%%%%%%%%%%%%%%
Similarly, in Figs.~3 and 4, the signal strengths observed at the CMS detector
and the CDF+D0 detectors, respectively, are shown.
%%%%%%%%%%%%%%%%%%%%%%%%%%%%%%%%%%%%%%%%%%%%%%%%%%%%%%%%%%%%%%%%%%%%%%%%%
\begin{figure} [h,t]                     %instead of \begin{figure}[t]
\begin{center}                        %instead of \begin{center}
\includegraphics[width=80mm]{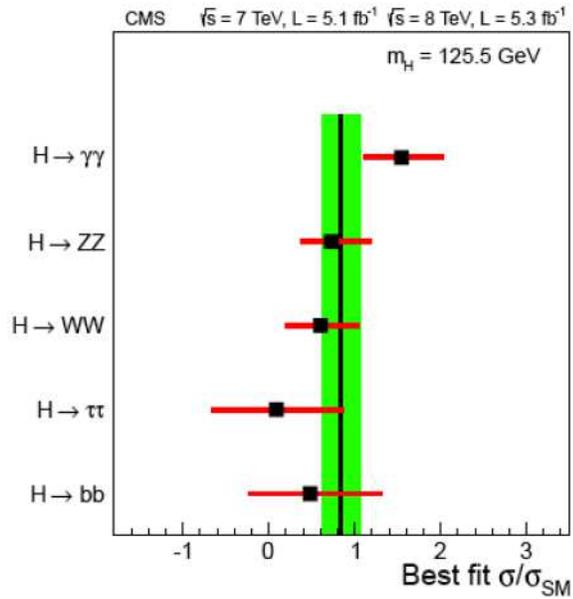}   % for LaTeX 2e
\end{center}                         %instead of \end{center}
\vspace{-2mm} \caption{The signal strengths of the 125-GeV boson decay channels
observed at the CMS detector~\cite{125GeVbosonCMS}.}
\end{figure}                        %instead of \end{figure}
%%%%%%%%%%%%%%%%%%%%%%%%%%%%%%%%%%%%%%%%%%%%%%%%%%%%%%%%%%%%%%%%%%%%%%%%%
%%%%%%%%%%%%%%%%%%%%%%%%%%%%%%%%%%%%%%%%%%%%%%%%%%%%%%%%%%%%%%%%%%%%%%%%%
\begin{figure} [h,t]                     %instead of \begin{figure}[t]
\begin{center}                        %instead of \begin{center}
\includegraphics[width=80mm]{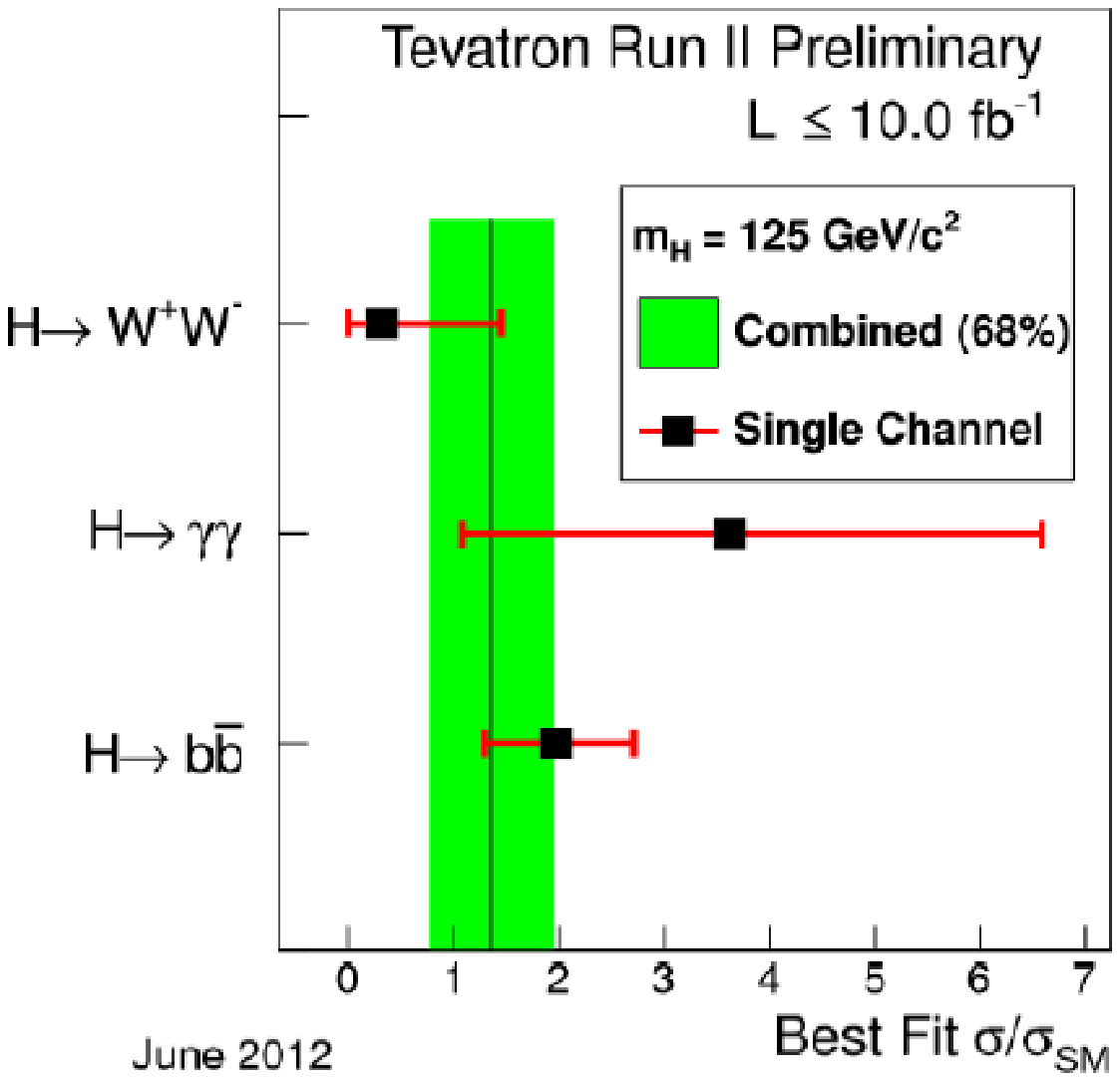}   % for LaTeX 2e
\end{center}                         %instead of \end{center}
\vspace{-2mm} \caption{The combined signal strengths of the 125-GeV boson decay channels
observed at the CDF and D0 detectors of Tevatron~\cite{125GeVbosonTevatron}.}
\end{figure}                        %instead of \end{figure}
%%%%%%%%%%%%%%%%%%%%%%%%%%%%%%%%%%%%%%%%%%%%%%%%%%%%%%%%%%%%%%%%%%%%%%%%%
Notice slightly stronger than SM-expected signal in the $h\rightarrow\gamma\gamma$
channel in both, ATLAS and CMS, graphs. While this observation might be inspiring
for theorists, it is not statistically strong enough to draw any serious conclusions.
Again, the full 2012 LHC data will probably shed more light into this question.

If the 125-GeV boson is the SM Higgs one there are no good reasons to expect
that the LHC will discover more new particles. If it is not the SM Higgs boson
new particles and forces are expected to exist. The question remains though 
whether they would be within the reach of the LHC.

%%%%%%%%%%%%%%%%%%%%%%%%%%%%%%%%%%%%%%%%%%%%%%%%%%%%%%%%%%%%%%%%%%%%%%%%%
\section{THEORY AFTER JULY, 4}
\label{sec:TheoryAfterJuly4}

The 125-GeV boson discovery along with various exclusion limits
derived from the 2011 and 2012 LHC data sets put a pressure on
the existing theories beyond the SM. The new findings have
forced the SUSY as well as Technicolor theories to begin
``organized retreat'': while SUSY theorists have to come to terms
with the absence of superpartner particles in quite a large range
of masses, Technicolorists have to deal with the fact of the existence
of the light boson of probably spin 0. Nevertheless, this is a very
healthy process promising a significant progress on the theory frontier,
the progress of the extent unmatched for decades.

Now, the interesting question is how the current LHC data enables to
discriminate among various candidates of beyond the SM (BSM) physics.
Aside from the (non)observation of new particles, the existing data
can be used to limit free parameters (like new couplings, energy scales, etc.)
of the particular BSM theories or of the Effective Lagrangians which would be
a more model-independent approach. 

There are BSM theories which predict $\mu\approx 1$. 
The SUSY as well as the strongly-interacting theories
can result in deviations of
the couplings of the 125-GeV boson from 
the SM Higgs boson values smaller than $10\%$.
This is illustrated in Table~2.
%%%%%%%%%%%%%%%%%%%%%%%%%%%%%%%%%%%%%%%%%%%%%%%%%%%%%%%%%%%%%%%%%%%%%%%%%
\begin{table}[ht]
\begin{center}
 \caption{Deviations of the couplings of the 125-GeV boson from 
          the SM Higgs boson values predicted by some BSM theories;
          $m_A$ stands for the mass of a heavy $A^0$ Higgs boson, and 
          $F$ is the Goldstone boson decay constant~\cite{Peskin}.}
\begin{tabular}{|c|c|c|}
\hline
theory & coupling & deviation \\
\hline\hline
 SUSY & $h\tau\tau$ & $10\%\cdot\left(\frac{400\;\mathrm{GeV}}{m_A}\right)^2$ \\
\hline
 SUSY (large $\beta$) & $hb\bar{b}$ & $\mathrm{dev}(h\tau\tau)+(1\leftrightarrow 3)\%$ \\
\hline
 composite Higgs & $hf\bar{f}$ & $(3\leftrightarrow 9)\%\cdot\left(\frac{1\;\mathrm{TeV}}{F}\right)^2$ \\
\hline
 Little Higgs & $hgg$ & $(5\leftrightarrow 9)\%$ \\
\hline
\end{tabular}
\end{center}
\end{table}
%%%%%%%%%%%%%%%%%%%%%%%%%%%%%%%%%%%%%%%%%%%%%%%%%%%%%%%%%%%%%%%%%%%%%%%%%

In~\cite{PeskinAccuracy}, M.~Peskin estimates the $h$ coupling measurement accuracy 
that can be achieved when the LHC collects
$300$~fb$^{-1}$ of data at the collision energy of $14$~TeV.
The estimates are shown in Fig.~5. Following the estimates it becomes clear 
that the LHC can
hardly become sensitive to the deviations cited in Table~2 even 
after reaching its full performance.
In Fig.~5 there are also the estimates of the $h$ coupling measurement accuracy 
achievable at various designs of future $e^+e^-$ colliders. 
This graph underlines the need for construction of such a collider
in order to measure the couplings of the 125-GeV boson with the precision
necessary to distinguish the individual BSM theories.
%%%%%%%%%%%%%%%%%%%%%%%%%%%%%%%%%%%%%%%%%%%%%%%%%%%%%%%%%%%%%%%%%%%%%%%%%
\begin{figure} [h,t]                     %instead of \begin{figure}[t]
\begin{center}                        %instead of \begin{center}
\includegraphics[width=80mm]{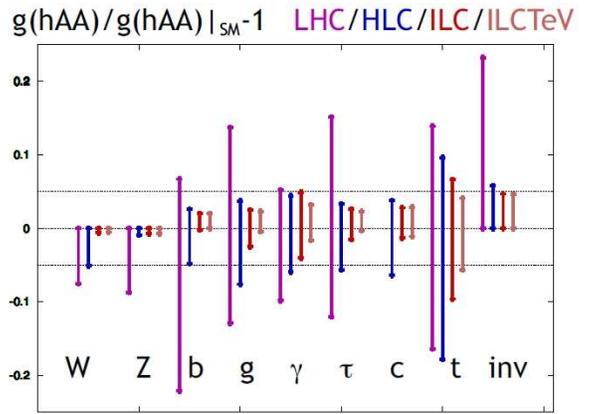}   % for LaTeX 2e
\end{center}                         %instead of \end{center}
\vspace{-2mm} \caption{The $h$ coupling measurement precisions 
                       achievable at the LHC($14$~TeV, $300$fb$^{-1}$)
                       and various designs of future $e^+e^-$ colliders:
                       HLC($250$~GeV, $250$~fb$^{-1}$),
                       ILC($500$~GeV, $500$~fb$^{-1}$), and
                       ILCTeV($1$~TeV, $1$~ab$^{-1}$),~\cite{PeskinAccuracy}.}
\end{figure}                        %instead of \end{figure}
%%%%%%%%%%%%%%%%%%%%%%%%%%%%%%%%%%%%%%%%%%%%%%%%%%%%%%%%%%%%%%%%%%%%%%%%%
Of course, we should keep in mind that if other new particles were discovered
at the LHC in the future it would provide complementary information and significantly alter 
the overall picture.

In a more model independent way, the electroweak symmetry breaking sector
with a scalar field $h$ can be parameterized by the non-linear effective
Lagrangian~\cite{EllisYou}
\[
 {\cal L}_{eff} \;\;=\;\; \frac{v^2}{4}\mbox{Tr}\left(D_\mu U D^\mu U^\dagger\right) \times
       \left[ 1+2a\frac{h}{v}+\ldots \right]
\]
\begin{equation}\label{eq:EffLagrangianScalar}
    -\left\{ \frac{v}{\sqrt{2}}\sum_f \bar{f}_L \lambda_f f_R
     \left[1+c_f\frac{h}{v}+\ldots \right]+\mbox{h.c.} \right\},
\end{equation}
where $U$ is a unitary $2\times 2$ matrix parameterizing the three 
unphysical Nambu-Goldstone bosons, $v$ is the conventional ESB scale,
and $\lambda_f$'s are the SM Yukawa couplings of the fermions $f$.
The coefficients $a$ and $c_f$ parameterize the deviations
of the $h$ couplings to massive electroweak gauge bosons and
to fermions, respectively, from those of the SM Higgs boson.
Note that the custodial symmetry is assumed in writing the Lagrangian.

In the following we will assume the flavor universality of the Yukawa
coupling deviations $c_f$: $c_t=c_b=c_c=c_\tau=\ldots\equiv c$.
The SM situation correspond to $a=c=1$. 

The global analysis of the available CMS, ATLAS, CDF and D0 data results
in the constraints on the deviations $a$ and $c$~\cite{EllisYou}. 
The constraints are shown in Fig.~6 along with the values of $a$ and $c$
predicted by some strongly-interacting BSM theories.
%%%%%%%%%%%%%%%%%%%%%%%%%%%%%%%%%%%%%%%%%%%%%%%%%%%%%%%%%%%%%%%%%%%%%%%%%
\begin{figure} [h,t]                     %instead of \begin{figure}[t]
\begin{center}                        %instead of \begin{center}
\includegraphics[width=80mm]{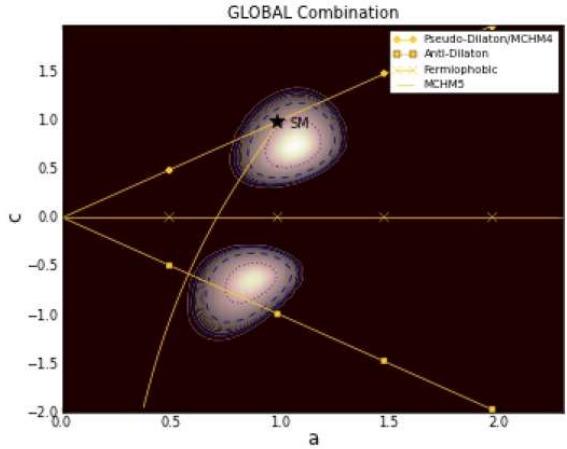}   % for LaTeX 2e
\end{center}                         %instead of \end{center}
\vspace{-2mm} \caption{The constraints on the deviations $a$ and $c$
                       of the 125-GeV boson couplings to the massive
                       electroweak gauge bosons and fermions, respectively.
                       The constraints have been obtained from the global
                       analysis of the CMS, ATLAS, CDF and D0 data in~\cite{EllisYou}.
                       The SM is represented by the black star, the yellow
                       lines represent various composite Higgs models.}
\end{figure}                        %instead of \end{figure}
%%%%%%%%%%%%%%%%%%%%%%%%%%%%%%%%%%%%%%%%%%%%%%%%%%%%%%%%%%%%%%%%%%%%%%%%%
As we can see the data excludes fermiophobic models ($c\rightarrow 0$)
while they admit pseudo-dilaton and MCHM4\footnote{The Minimal Composite Higgs Model 
embedded into spinorial representation of $SO(5)$.} models with parameters close
to the SM model case.

Other global fit has been performed in~\cite{EllisYou} testing the possibility that
$h$ couples to other particles proportionally to some powers of their masses.
In the SM,
\begin{equation}
 \lambda_f=\sqrt{2}\frac{m_f}{v},\;\;\;
 g_V = 2 \frac{M_V}{v},
\end{equation}
where $g_V$ and $M_V$ are the gauge coupling and the mass of the massive
gauge boson $V$. The coupling-mass linear relations can be generalized to
the following anomalous scaling laws
\begin{equation}
 \lambda_f=\sqrt{2}\left(\frac{m_f}{M}\right)^{1+\epsilon},\;\;\;
 g_V = 2 \left(\frac{M}{v}\right)^{1+2\epsilon},
\end{equation}
which in terms of the effective Lagrangian~(\ref{eq:EffLagrangianScalar})
translates to the deviations
\begin{equation}
  a = \frac{v}{M}\;\left(\frac{M_V}{M}\right)^{2\epsilon},\;\;\;
  c_f = \frac{v}{M}\;\left(\frac{m_f}{M}\right)^{\epsilon},
\end{equation}
where $M$ is a free parameter of the mass dimension.
The SM is being recovered when $\epsilon=0$ and $M=v$.
The mass independent scenario would require $\epsilon=-1$.

The constraints on the parameters $M$ and $\epsilon$ obtained~\cite{EllisYou}
from the global fit of the CMS, ATLAS, CDF and D0 data are shown in Fig.~7.
%%%%%%%%%%%%%%%%%%%%%%%%%%%%%%%%%%%%%%%%%%%%%%%%%%%%%%%%%%%%%%%%%%%%%%%%%
\begin{figure} [h,t]                     %instead of \begin{figure}[t]
\begin{center}                        %instead of \begin{center}
\includegraphics[width=80mm]{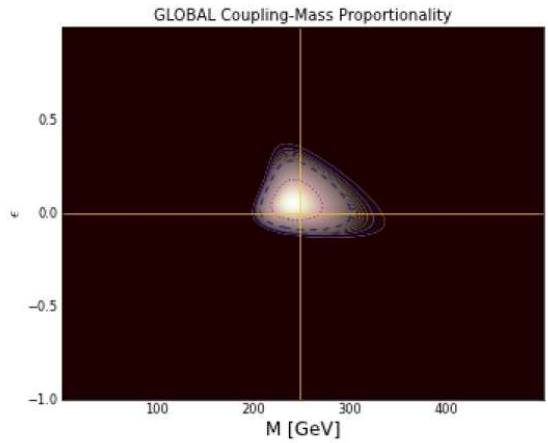}   % for LaTeX 2e
\end{center}                         %instead of \end{center}
\vspace{-2mm} \caption{The constraints on the scaling parameters $M$ and $\epsilon$
                       of the 125-GeV boson.
                       The constraints have been obtained from the global
                       analysis of the CMS, ATLAS, CDF and D0 data in~\cite{EllisYou}.
                       The SM corresponds to the intersection of the yellow lines.}
\end{figure}                        %instead of \end{figure}
%%%%%%%%%%%%%%%%%%%%%%%%%%%%%%%%%%%%%%%%%%%%%%%%%%%%%%%%%%%%%%%%%%%%%%%%%
Clearly, the preferred values of the parameters are consistent with the SM
Higgs boson scenario
\begin{equation}
 \epsilon = 0.05\pm 0.08,\;\;\;
 M = 241\pm 18\;\mathrm{GeV}.
\end{equation}

Even though the current data are consistent with the 125-GeV SM Higgs boson
alternative possibilities remain well open. Fitting the alternative theories
to the existing data to recognize survivors is the task of the utmost importance.
At the same time, equally important is the experimental search for new particles.
Any such discovery (or a new exclusion limit) will provide a supplementary
information the value of which cannot be overestimated. For both these tasks
the formalism of the effective Lagrangians is a very useful tool as it was illustrated
above by the parameter analysis based on the effective Lagrangian~(\ref{eq:EffLagrangianScalar}).

%%%%%%%%%%%%%%%%%%%%%%%%%%%%%%%%%%%%%%%%%%%%%%%%%%%%%%%%%%%%%%%%%%%%%%%%%
\section{THE 125-GEV BOSON AND THE TOP-BESS MODEL}
\label{sec:tBESSModel}

The strongly-interacting BSM theories typically introduce new strong interactions
and new elementary fields which are subject to them. 
Thus, it is reasonable to expect that the new fields will form the bound states
of various masses, spins, etc., as in the QCD. And as it the case in the QCD, 
the new strong interactions are not treatable perturbatively. Following the QCD
experience, physics of the new bound states can be described by the effective
Lagrangian formulated in terms of the bound state fields and obeying the new
physics symmetries.

We have formulated the top-BESS model~\cite{topBESS} as a modification of the BESS 
model~\cite{BESS}. The BESS model is an example of the effective Lagrangian
describing the Higgsless ESB sector with an extra $SU(2)$ 
vector bound state triplet. Recently, the model has been utilized
in the context of the extra-dimensional 
theories~\cite{3siteExtraDimBESS,4siteExtraDimDBESS}.

Both, the BESS and top-BESS models, have been formulated without 
scalar resonances of any kind because their primary motivation 
was the systematic effective description and study of a vector bound state
physics. Nevertheless, the introduction of a scalar resonance 
to the BESS-like effective Lagrangian is not a difficult task; 
actually, it is much less involved than in the case of the vector resonance.

Both, BESS and top-BESS, Lagrangians possess the same symmetry. Their 
global symmetry $SU(2)_L\times SU(2)_R\times U(1)_{B-L}\times SU(2)_{HLS}$
is spontaneously broken down to $SU(2)_{L+R}\times U(1)_{B-L}$ while 
the local symmetry $SU(2)_L\times U(1)_Y\times SU(2)_{HLS}$ is spontaneously
broken down to $U(1)_{em}$. The electroweak gauge symmetry is enlarged
by the auxiliary $SU(2)_{HLS}$ gauge group where ``HLS'' stands for 
the Hidden Local Symmetry. In the BESS model, the new vector triplet is
introduced as a gauge field of $SU(2)_{HLS}$ with the gauge coupling $g''$.

In the top-BESS model we modify the direct interactions of
the vector triplet with fermions. While in the BESS model there is
a universal direct coupling of the triplet to all fermions of a
given chirality, in our modification we admit direct couplings of
the new triplet to top and bottom quarks only. Our modification is
inspired by the speculations about a special role of the top quark
(or the third quark generation) in the mechanism of ESB. 
The speculations are fueled by the observation that the 
large top mass is surprisingly close to
the ESB scale: $m_t\approx v/\sqrt{2}$.

%%%%%%%%%%%%%%%%%%%%%%%%%%%%%%%%%%%%%%%%%%%%%%%%%%%%%%%%%%%%%%%%%%%%%%%%%
\begin{figure} [h,t]                     %instead of \begin{figure}[t]
\begin{center}                        %instead of \begin{center}
\includegraphics[width=80mm]{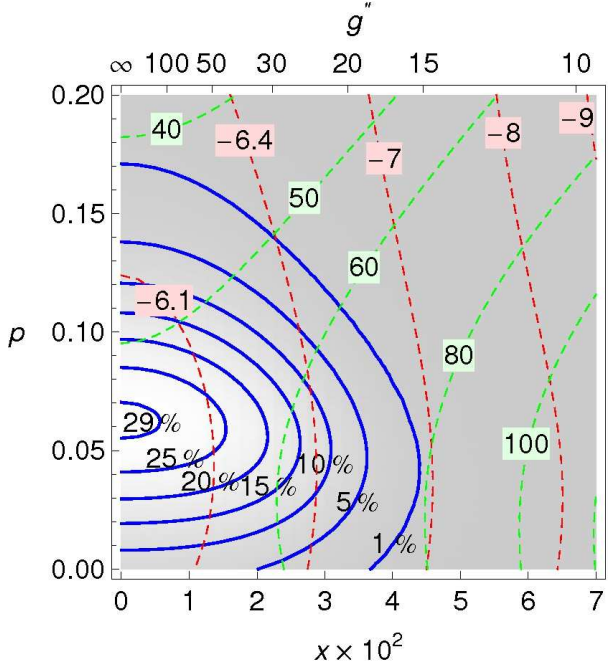}   % for LaTeX 2e
\end{center}                         %instead of \end{center}
\vspace{-2mm} \caption{The most preferred values of the $(p,g'')$ parameters
(white area)
when there is no real scalar field in the top-BESS Lagrangian. 
The blue iso-lines join $(p,g'')$ points with the same statistical 
support; the values of the support are shown in percents placed
next to the lines.}
\end{figure}                        %instead of \end{figure}
%%%%%%%%%%%%%%%%%%%%%%%%%%%%%%%%%%%%%%%%%%%%%%%%%%%%%%%%%%%%%%%%%%%%%%%%%
%%%%%%%%%%%%%%%%%%%%%%%%%%%%%%%%%%%%%%%%%%%%%%%%%%%%%%%%%%%%%%%%%%%%%%%%%
\begin{figure} [h,t]                     %instead of \begin{figure}[t]
\begin{center}                        %instead of \begin{center}
\includegraphics[width=80mm]{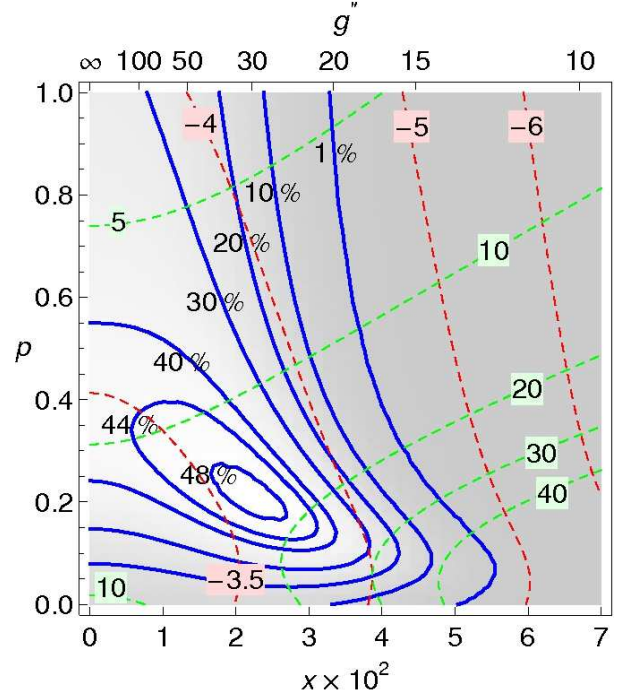}   % for LaTeX 2e
\end{center}                         %instead of \end{center}
\vspace{-2mm} \caption{The most preferred values of the $(p,g'')$ parameters
(white area)
with the 125-GeV scalar field in the top-BESS Lagrangian. 
The blue iso-lines join $(p,g'')$ points with the same statistical 
support; the values of the support are shown in percents placed
next to the lines.}
\end{figure}                        %instead of \end{figure}
%%%%%%%%%%%%%%%%%%%%%%%%%%%%%%%%%%%%%%%%%%%%%%%%%%%%%%%%%%%%%%%%%%%%%%%%%

In the top-BESS model, we take the possible chirality dependence of
the triplet-to-top/bottom coupling into account multiplying the
$SU(2)_{HLS}$ gauge coupling $g''$ by the $b_L$ and $b_R$
parameters for the left and right fermion doublets, respectively.
In addition, we can disentangle the triplet-to-top-quark right
coupling from the triplet-to-bottom-quark right coupling. This
breaks the $SU(2)_R$ symmetry which is broken by the SM
interactions, anyway. For the sake, we have introduced a free
parameter, $0\leq p\leq 1$. The $p$ parameter can weaken the
strength of the triplet-to-b$_R$ coupling. However, the $SU(2)_L$
symmetry does not allow us to do the same splitting for the left
quark doublet.

We have performed a multi-observable fit of the top-BESS parameters~\cite{LEtopBESS}
using (pseudo)observables $\epsilon_1$, $\epsilon_2$, $\epsilon_3$,
$\Gamma(Z\rightarrow b\bar{b})$, and BR$(B\rightarrow X_s\gamma)$.
The epsilons are related to 
the \textit{basic observables}~\cite{EpsilonMethod}:
the ratio of the electroweak gauge boson masses, $r_M\equiv M_W/M_Z$,
the inclusive partial decay width of $Z$ to the charged leptons,
$\Gamma_\ell(Z\rightarrow\ell\bar{\ell}+\mathrm{photons})$,
the forward-backward asymmetry of charged leptons at the $Z$-pole,
$A_\ell^{FB}(M_Z)$, and
the inclusive partial decay width of $Z$ to bottom quarks,
$\Gamma_b(Z\rightarrow b\bar{b}+X)$.

In Fig.~8 we show the most preferred values of the $(p,g'')$ parameters
when there is no real scalar field in the top-BESS Lagrangian. 
We can see that data pushes $g''$ to infinity.
In the limit $g''\rightarrow\infty$ the influence of the vector triplet
on low-energy physics disappears, though. 

In Fig.~9 we can see the impact of adding the 125-GeV scalar particle
into the top-BESS Lagrangian. The addition is performed by extending
the top-BESS Lagrangian with the Lagrangian~(\ref{eq:EffLagrangianScalar}).
We have calculated the preferred $(p,g'')$ values in the limit $a=c=1$.
Then, the best value of $g''$ moved to about $40$ and the best value of
$p$ is close to $0.2$. Also, the statistical backing for the best values
has grown. Hence, the inclusion of the 125-GeV scalar into the top-BESS
model improves its chance that it might correspond to the actual 
strongly-interacting situation
in nature. Data suggests that should it be the case the new vector triplet
would interact with the right bottom quark some five times weaker than
with the right top quark. 

%%%%%%%%%%%%%%%%%%%%%%%%%%%%%%%%%%%%%%%%%%%%%%%%%%%%%%%%%%%%%%%%%%%%%%%%%
\section{CONCLUSIONS}

On July, 2012, the new Higgs era in high energy physics
started by discovering the 125-GeV boson. While we are not yet sure 
if the boson is the SM Higgs boson the LHC data resembles it. To settle down
this question more LHC data will be needed. Yet it might be the case
that even all LHC data will not be enough to solve this question.
Particularly, if there will be no more new particles discovered 
at the LHC, new $e^+e^-$ colliders will be needed to measure
the 125-GeV boson couplings with sufficient precision.

Major BSM scenarios, like the SUSY and Technicolor, 
are not excluded by the current LHC data. Nevertheless, in both camps 
many representatives have not survived the LHC findings.
And the process continues as we speak.
The full 2012 LHC data can be full of surprises or reveal nothing new.
It is really exciting time for high energy physics!

\noindent ACKNOWLEDGMENT: I would like to thank the organizers 
of the 19th Conference of Slovak Physicists for inviting me to
present a talk on such an exciting topic.

\end{document}